\documentstyle[12pt,aasms4]{article}
\begin{document}

\title{Environments of QSOs at redshift 0.9 to 1.3\footnote{Based on
observations obtained with the Canada France Hawaii Telescope, which
is operated by CNRS of France, NRC of Canada, and the University of 
Hawaii}}

\author{J.B. Hutchings, P. Scholz }

\affil{Herzberg Institute of Astrophysics, 5071 West Saanich Rd.,
Victoria, B.C. V9E 2E7, Canada; john.hutchings@nrc.ca}

\author{L.Bianchi}

\affil{Dept of Physics and Astronomy, Johns Hopkins University, 
3400 N. Charles St., Baltimore, MD21218}

\begin{abstract}
We analyse new deep g and i-band imaging with the CFHT of 16 QSOs in the 
redshift range 0.9 to 1.3. The principal points of interest are the 
symmetry and signs of tidal effects in the QSO hosts and nearby 
(`companion') galaxies. The sample measures are compared
with similar measures on randomly selected field galaxy samples. Asymmetry
measures are made for all objects to g $\sim$22, and magnitudes of all
galaxies 2 magnitudes fainter. The QSOs are found in denser environments
than the field, and are somewhat offset from the centroid of their 
surrounding galaxies. The QSO hosts appear more disturbed than other 
galaxies. While the QSO companions and field galaxies have the same
average asymmetry, the distribution of asymmetry values is different. 
QSO companions within 15 arcsec are fainter than average field galaxies. 
We discuss scenarios that are consistent with these and other measured
quantities.
\end{abstract}

\section{Introduction and observations}

Imaging studies of QSOs have ranged from the local universe to those at 
redshift near to 6. From these and related studies, it is widely accepted 
that nuclear activity arises from accretion episodes on to massive central
black holes in galaxies, and there are many scenarios for the process.
Given the evidence for the connections between galaxy stellar properties
and the central black hole mass, there is interest in studying QSO
episodes in galaxies, at all redshifts, for clues as to how the black
hole and galaxies develop this connection (e.g. Salviander et al 2007). 
At low redshifts, there is
considerable evidence that QSO host galaxies have had recent merging
of tidal disturbance events, which activate the nuclear accretion process.
At redshifts 2 and higher, host galaxies appear to be in their early stages
of assembly, with high star-formation. Thus, these probably represent
different stages in the evolution of both galaxy and central black hole.

There are few environmental studies of QSOs at redshifts near 1.
Falomo et al (2004) and Kotilainen et al (2007) discuss samples at
redshift 1.2 and higher, while Kukula et al (2001) investigated
a sample of 9, with redshift 0.8 to 1.0. There are several
investigations at redshifts 2 and somewhat higher (e.g. Ridgway et al 
(2001), Hutchings et al (2002)). Most of these investigations were
done at NIR wavelengths, where the contrast with the nucleus is
expected to be better, and where ground-based AO works best. The main
focus of these papers is the nature of the host galaxy, and there
is some consensus that radio-loud objects have higher luminosity
hosts, but many are luminous elliptical-type galaxies. Lacy (2006)
gives a good summary of the work on higher redshift QSOs to that date.

The redshift range around 1 is when large galaxies are assembled, and may 
be in the process of forming the disk structures common in the present epoch. 
This work describes an investigation of QSO host galaxies and environments 
to learn more about this stage of their evolution, based on a new and 
relatively bias-free sample. Our resolution and wavelength range are 
not optimised for host galaxy details, and the main focus of this paper is
galaxies that may be associated with the QSOs, and their morphology. 
In particular, we are interested
in the asymmetry of such galaxies (and QSO hosts), which can be characterised
by deviations from elliptical contours with radius, even if small
scale details are not well resolved. We are particularly interested
in asymmetry of the fainter outer parts of galaxies, which arise from
recent merging and tidal events among them. Few QSO studies in this
redshift range have discussed this aspect. Wold et al (2001) discuss
the galaxy environments of QSOs at redshift 0.5 to 0.8 in a general
way. Hutchings and Proulx (2008) discuss companion galaxy asymmetries 
in a low redshift sample, and this paper is a similar investigation
near redshift 1.  

The sample is derived from the combined SDSS and GALEX surveys, which
overlap in the shallow and medium GALEX surveys (see e.g. Bianchi et al 2007).
Initial samples of some 3000 and 14000 objects, respectively, were drawn from
two-colour planes such as FUV-NUV vs NUV-r, in the locations where QSOs
separate out from stars and galaxies (see Bianchi et al 2005; Bianchi 2008).
Within these is
a subsample of about 700 objects identified as QSOs by their SDSS spectra.
We selected those in the redshift range 0.9 to 1.3, and declination
10$^o$ to 40$^o$, for imaging with CFHT. Observations were made of 16 of 
these with the Megaprime camera of the CFHT. This sample is thus a subset of
those investigated by Bianchi et al (2007), which have photometric
coverage from 1500 to 9000\AA. 

   The observations consist of images with g and i filters, of exposure
400 and 500 seconds, respectively, carried out between October 2006
and January 2007. At least two exposures were taken in each filter, and
several had more. Image quality was in the range 0.6 to 1.2 arcsec FWHM for
most of the observations. Table 1 summarises the QSOs observed. 
The QSO absolute magnitudes are in the range -24 to -26 in g-band,
which is in the middle of the range for QSO luminosities. 
At the redshift of the QSOs, the observations are sampling rest band 
wavelengths about 2400 and 3800 A. The FIRST radio catalogue does not
cover the part of the sky where the sample is, so we do not know
if any are radio-loud. 

\section{Measurements}

  The QSOs were first identified on each CFHT image. They are located
at the same spot among the 36 CCDs, about 12 arcsec from the edge of one.
This means there is a gap of about 11 arcsec of sky on this side of
each QSO. We identified and made measurements of all galaxies within
75 arcsec (400 pixels) of the QSO, with exceptions where they lie close
to bright stars
or image flaws.  Several stars were also selected and measured to
characterise the PSFs in each image. For each object and the QSO itself,
the flux and position were recorded using the `imexamine' task of IRAF.

  For all except the faintest galaxies, the `ellipse' task was used to fit
contours at a fixed set of radii increasing uniformly on a log scale.
For each contour, an asymmetry index
is calculated as (Contour level)$^{1.5}$ x (sum of the absolute values of
the third harmonic deviations from an ellipse) x (semi-major axis)$^{1.5}$.
This quantity is slightly different (has a different signal level
exponent) from that used by Hutchings and
Proulx (2008), but is less dependent on the total signal so that objects
of widely different flux as well as size can be compared. As in 
Hutchings and Proulx, the weighting also
is designed to be sensitive to faint outer asymmetries that are
signatures of tidal events, which otherwise would be overwhelmed by
small asymmetries (such as dust lanes or disk arms or nuclear saturation)
seen in the bright inner contours. Plots of this index with contour radius
show where the asymmetries lie within each image. A mean asymmetry
index is the mean of the individual contour asymmetry values lying between
radii 4 pixels (about the resolution limit of the observations),
and where the signal lies less than 10\% above the sky value.
This mean is normalised by the total signal from the object,
so that galaxies of different brightness can be compared. Hutchings
and Proulx discuss other tests and properties of this index.

  The images of the faintest galaxies were too noisy for ellipse-fitting,
but we have recorded their position and flux, down to 3000 counts in each
image. We used the SDSS magnitudes from Table 1 to calibrate the 
images, and this limit corresponds to g$<$24.1 and i$<$24.4.  
The asymmetry measures were derived for galaxies to about 1.5 magnitudes 
brighter than these values. Galaxies overall brighter than the related 
QSOs are considered to be foreground objects and were not
measured.  In general, all individual exposure images were measured, as well 
as combined images, for each filter and for both filters combined. Combined
images were constructed by shifting and adding, so as to eliminate flaws
and cosmic ray events, as well as increase signal to noise.
In a few cases, single pixel cosmic rays were edited out by hand, where 
they were affecting the asymmetry measurements. In all cases, accurate 
sky subtraction was
important, and great care was taken with this. Fluxes determined by
`imexam' and `ellipse' were compared, since they involve independent
estimates of the sky level, and had to agree within 10\% for the
flux to be adopted as final. In the case of faint objects that
were not fitted with ellipse, both imexam and subarray pixel statistics
around the objects were used for this comparison. Where this agreement was 
not obtained, the sky levels were adjusted until they did.

  We made all the same measures in control fields, for comparison with the
QSO fields. These were centred on positions within the same CCD as the
QSO, lying more than 2arcmin from the QSO and at least 400 pixels of the
CCD edge. The positions within this roughly 4 x 9 arcmin area of sky,
were selected by a random number generator. Where 
there were multiple exposures of QSOs, the same control objects were used.
Table 2 gives average asymmetry values for different object classes. 
We discuss details in subsequent sections. 

  Figure 1 shows the distribution of image quality, by star FWHM values.
The non-integral bin values arise from using pixels as the units of measure. 
The figure also shows the values for the QSOs. While very similar, the
QSO distribution is skewed to higher values, which suggests they are
marginally extended compared with the stars, although the formal comparison
of the distributions shows only 20\% probability they are different.

\section{Biases and galaxy counts}

    Before looking for trends and differences between subsamples,
we need to look at possible biases in the dataset. The obvious ones to
check are image quality, photometric calibration, and dependence on
redshift and magnitude of objects.

   Figure 2 shows the photometric calibration, based on the SDSS g and i
magnitudes, and the mean fluxes for them from our data.  Zero points
have been applied to each colour in the plots. The scatter increases for 
fainter ojects, but in QSOs there is always some possibility of real 
variability
at the level of a few tenths of a magnitude, betwen the observation epochs.

  Figure 3 shows some illustrative plots. They show there is little or no
dependence of the asymmetry value with redshift, signal level, or
image quality. A few of the stars have high asymmetries in good image
quality, which we ignore as due to flaws or faint blends.
The lowest asymmetry values are not seen in the faintest
stars, but these values are lower than any seen in QSOs or galaxies,
and the value is robust over the factor 100 signal range of all QSOs
and galaxies in our program. The redshift range is relatively small
and no trend is seen among the individual QSO measures.

  Because the QSOs were situated close to a CCD edge, and hence a gap in
sky coverage, it is necessary to correct the galaxy counts for the missing
area of sky. The gap amounts to approximately 10\% of the total area
investigated. The way we made the correction was to reflect the galaxies
an equal distance from the QSO on the other side of the gap, into the
gap, and use these for statistics on galaxy properties with distance from
the QSO.  Figure 4 shows the stacked distribution of all the galaxies
measured, for all QSO and all control fields. In the QSO distribution,
the gap lies between +12 and +23 arcsec from the QSO.

  Figure 5 shows the
sky density of galaxies around the QSOs and in the control fields.
The lower plots are galaxies with asymmetry measures and the upper plots
include all the fainter galaxies to the limits mentioned above. For an
even sky density, the plots should form straight lines. The dotted
lines are those corresponding to the control fields. It appears that
the QSOs lie in regions of greater galaxy density, especially in the
radius range 30 to 50 arcsec (150 to 270 pixels). Both sets appear
to suffer from some incompleteness in the largest radius bin. Within the
areas covered, the QSO fields have 28\% more galaxies, and within 1 arcmin
they are 17\% higher. The QSO measurable companion galaxies are 14\%
more numerous within 1 arcmin.  The QSO fields thus overall have an 
excess of order 25\%
over the foreground and background galaxy counts in our sample. This needs
to be taken into account in looking for differences between the QSO and field
galaxy environments.

   The ratio of total galaxies down to our faint flux limit, to 
ellipse-measured galaxies is close to 3 (see Figure 5), but with
considerable scatter. Galaxy counts show trends with QSO redshift, varying
by a factor 2 over the redshift range 0.9 to 1.3, with higher
total counts in the lower redshift cases. While there is considerable
scatter, this suggests that there are real groups of galaxies associated
with the QSOs, and we are seeing a flux limit effect with redshift. 
However, the scatter does not warrant any more detailed conclusion.

   It is of interest to see if the QSOs are centrally placed with respect
to their surrounding galaxies. The offset between mean position of all
surrounding galaxies, and the QSO, should become smaller with increasing
projected distance from the QSO, if their positions are random and
unconnected with the QSO. If the QSO is centrally placed within a real 
group, this offset will be small for all projected distances.
These offsets for the control fields do indeed decrease with increasing 
sky radius, as expected for random control field origins. The QSO 
offsets however, become systematically larger with increasing radius, to
our radius limit of just over 1 arcmin. This is consistent with an 
associated set of real companions, among which the QSOs are offset 
by an average of about 5 arcsec. The overall mean offset of companions for 
all QSOs, is the same as for the control groups, which is consistent
with the individual QSO offsets being in random directions, amongst 
our sample fields.

\section{Photometry}

   Figure 6 shows the magnitudes of galaxies in approximately 10 arcsec
annuli around the QSOs, compared with the same for the randomly placed
control positions. We have plotted both mean and median values, as they
differ somewhat due to the non-gaussian distribution of magnitudes.
The magnitudes themselves are from the combined g and i-band images,
for maximum signal on the fainter ones. The standard deviations of the
means are typically 1 magnitude, but for the innermost radius
bins, they are about 0.5 mag. As expected, the magnitudes
in the control fields show no trend with position. However, the QSO
fields have fainter galaxies in the inner bins, and brighter galaxies
in the middle-range bins, which is where the galaxy count excess is found.
The suggestion is that the QSO-associated galaxies are faint by some
0.5 magnitudes near to the QSO and the excess galaxies at 30-50 arcsec
radii are brighter than field galaxies by a few tenths of a magnitude.

   We may compare the g-i colours of galaxies from our data. The control
field galaxies and the QSO companions have mean and median g-i colours
close to 0.93 for both groups. The companions in the 25-50 arcsec distance
range also have this colour, so there are no large mean colour differences
between these populations. But the QSO companions have fewer blue galaxies 
and fewer red galaxies than the field - the colour range is much less 
spread, and formally the two distributions are different at the 96\% 
level. This indicates a more uniform
population, among the QSO companions, with none in star-formation and
fewer old galaxies. Bearing in mind that $\sim$80\% of the galaxies 
are foreground or background (from the 25\% excess in the QSO fields, as
noted in the previous section), 
this indicates a very different population among the true QSO companions.
The QSO colours are much bluer, with an average value of 0.34.

    Figure 7 shows the 7 filter magnitudes from GALEX and SDSS, 
for all the sample.  The Lyman
break lies between the FUV and NUV for this redshift range, and Ly$\alpha$
lies in the NUV band. The distribution is much the same for all
individual objects in the sample. We compare these with similar average
magnitudes for normal QSOs at other redshift (Bianchi et al 2008),
with the bands shifted to bring them all to redshift about 1.1. 
While the filter bandpasses do not match exactly the same rest wavelengths 
in these comparisons, there appears to be little difference between them.
The low redshift sample has lower luminosity, and hence presumably
lower Ly$\alpha$ flux, and possibly some host galaxy light, which 
would account for the relatively fainter NUV and u magnitudes for these. 
The sample in this paper does not show the extreme UV characteristics
discussed by Bianchi et al (2008). The SDSS spectra cover the rest 
wavelength range from C III] 1909 to H$\gamma$, and are not
significantly different from the spread of SDSS spectra for randomly
selected QSOs in this redshift range.  

    Looking in more detail at the photometry, the NUV-i index shows a trend
of increasing with redshift, with two exceptions. The exceptions make the
overall trend non-significant, but it is what is expected as the Lyman
absorption
moves into the NUV bandpass. The FUV-NUV index shows a lot of scatter and
no significant trend with redshift. The FUV and NUV magnitudes show no
correlations with redshift, and do correlate well with the SDSS magnitudes.
This shows up in individual plots of the format of Figure 7, as most
lines connecting dots for individual QSOs do not cross each other. 
There are nominal values of 
foreground reddening based on the lines of sight from the extinction
maps of Schlegel et al (1998), but the values are
small and show no correlation with FUV magnitude or g-r index. Thus
we expect that extinction plays no significant role in the sample.

\section{Asymmetry in galaxies}

  Individual mean asymmentry values on the same object agree to within
10\%, with an average spread about half this. The i-band images show higher
asymmetry than the g-band. For QSOs the asymmetry in i-band is 29\%
higher than in g-band (which drops to 15\% if we ignore 3 high outliers). 
For galaxies the i-band asymmetry excess is 14\%, while for stars it 
is only 5\%. Thus, while the i-band images may be slightly 
more intrinsically asymmetric, there seems to be a real difference in the
galaxies, in the same sense. In addition, the mean asymmetry (g and i
combined) is higher for blue galaxies - both QSO companions and field 
galaxies. The median asymmetry rises by 50\% as g-i goes from 1.75 to 0.25 and
the mean value rises by 32\%, for all galaxies. Galaxies that are blue are
more asymmetric, but the asymmetry is more apparent in redder light. 
This may arise from a combination of star-formation and dust, for instance.
There is no significant change with UV-optical colour: this also may be
a result of dust, which has greater extinction in the UV. As a `sanity 
check' we noted galaxies that appeared on inspection to be interacting 
or disturbed. These galaxies on average had measured asymmetry 
indices 2-3 times that for the others. 

  Figure 8 shows the mean values of the asymmetry indices for
the different classes of object. The first point to note is that the
QSO companion galaxies and the field galaxy sample have essentially 
the same mean 
asymmetry. Thus, there is no gross difference in morphology. The
value for stars is lower, confirming that we are resolving and measuring
the galaxy morphology. Star values contain some high outliers, probably
due to faint galaxies or structures within the PSF, so we have derived 
a median value
for bright stars, matching the QSO magnitudes, for comparison with the
QSO asymmetries. The distributions of asymmetry values for galaxies and
QSOs are well spread about their means and their median values are 
essentially the same. The distribution of asymmetry values for QSOs
and matched magnitude stars are different at the 98\% level, by the
S-Z test, so that the QSO asymmetries are not simply the effect of the
nuclear point spread function. 

Thus, while the QSOs are dominated by their unresolved nuclei, and have 
low asymmetry values, as seen in Table 2, we can attempt to estimate 
the asymmetries of the host galaxies. The average absolute magnitude of 
the QSO nuclei is a factor 20 brighter than L*, so to estimate their host 
galaxy asymmetries, we have renormalised the QSO values by a (conservative)
constant factor 10, after subtracting the median asymmetry for bright 
stars (0.005), which is the lowest asymmetry that can apply to any of 
our objects. The QSO magnitudes are increased in Figure 8 by the same 
factor 10 (2.5 mag) to approximate the host galaxy magnitudes.  
The host galaxy asymmetries, derived this way, are higher than the 
non-host galaxies (Fig 8), and this signal presumably lies in 
the faint outer parts of the images. As an independent approach, we
note the FWHM values for the QSO images are on average 2\% larger 
than the stars, with a total spread from 97\% to 109\% of
the star values from the same images.  

Figure 9 shows the asymmetry as a function of distance from the QSO
(and random sky points for field galaxies). There is no obvious trend
with projected distance from the QSO, or, as already noted in Figure 8,
overall difference from the field galaxies. However, we have sketched in
lines that suggest a higher asymmetry population, and a lower asymmetry
population for the QSO companions, and a single dashed line for the field
galaxies. Bearing in mind that there are projection effects and line of
sight contaminants, the plot is consistent with companion galaxies that
have higher asymmetries near the QSO, and perhaps also some that have lower
than normal asymmetry. From Figure 8, the low asymmetry galaxies are
faint. Possibly we are seeing galaxies that are disturbed and some that 
have been stripped, as a result of being in a group environment. 

   Figure 10 shows the distributions of QSO and field galaxy asymmetry values,
of combined g and i images. The field galaxies are overall skewed to higher
values than the QSO galaxies, except for the highest values. This difference 
is more marked in i than g-band data. A Kolmogorov-Smirnov test indicates 
the distributions in Figure 10 are different at the 98\% confidence level. 
In all sets of filter images the difference is significant at over 90\%.  
If we look only at the galaxies within 35arcsec of the QSO, the
double-peaked distribution of the QSO companion asymmetries is more
marked, as can be seen in Figure 9. The K-S test actually gives
lower probability for this kind of difference, since the field galaxies 
have a single-peaked broader distribution of asymmetries. We note that
the faintest galaxies have no asymmetry measures, so we are unable to
examine this property of the faint galaxies that lie closest to the QSOs
(see Figure 6).

\section{Summary and discussion}

   The sample of QSOs we have investigated appear to be normal in their
luminosities and spectral energy distributions, and lie in the redshift
range 0.9 to 1.3, which is not well investigated for QSO environments. 
The dataset is uniform and has average resolution 0.9 arcsec. The
QSOs are marginally resolved with the data. 

The photometric
calibration is good to a few tenths of a magnitude, and there are no
systematics with image quality, magnitude, and QSO redshift within the ranges
present. Control fields in the same data have been used extensively to
enable good comparison with the QSO environments. 

   The QSOs are on average not centrally located with their surrounding
galaxies, down to about 24 magnitude. The QSOs appear to live in 
significantly denser environments than random places in the sky. 
The QSO companion galaxies have significantly different colour
distribution from the field, having less spread in colour. 
The QSO environments have significantly fainter galaxies within
$\sim$15 arcsec, than the field. The galaxies investigated in the
QSO environments are estimated to contain about 75\% non-associated
line of sight galaxies in the range of brightness measured. 

   Making an average correction for the nuclear point source, the host
galaxies are about two times more asymmetric than the companion or field 
galaxies in the same range of luminosity. The companion and field galaxies
have about twice the asymmetry as stars in the fields. The companion galaxies
do not have the same asymmetry distribution as field galaxies, having
more symmetrical and more asymmetrical ones, in a double-peaked distribution.
The more asymmetrical galaxies lie within 30 arcsec of the QSO.

   The picture that emerges for this sample is that the QSOs have
asymmetrical structure, as do their closer bright companion galaxies 
in compact groups of diameter some 600Kpc. In the innermost 100Kpc the
companion galaxies are fainter than the average field galaxies, and
there are also several galaxies with higher than normal symmetry.

  This is consistent with interactions within the groups, some of which 
triggered the QSO event. The number of faint and symmetrical galaxies might
indicate a population of post-interaction stripped galaxies with little
star-formation. The brighter asymmetrical galaxies are bluer and have
higher asymmetry, indicating ongoing or recent interation events. 

   The excess galaxies in the QSO fields (See Figure 5) are 1.9
within 90Kpc, 3.8 within 280 Kpc, and 7.5 within 365 Kpc, average per QSO.
These numbers are similar to the excess galaxies found at z$\sim$0.3
by Hutchings and Proulx (2008). However, the absolute magnitude of the
galaxies in this sample have a faint limit about one magnitude more
luminous than those in the low redshift sample. K-corrections may
reduce that, but this is still a high density of galaxies, in which
interactions must be common. We note that the excess is not seen within
100 Kpc of the QSO, so it may be that the QSO is clearing out the central
part of the group, perhaps by merging.  

   Our data indicate that QSOs at redshift near 1 live in dense small
groups of galaxies, in which merging and tidal interaction are
occurring frequently, and that the QSOs themselves are affecting
the galaxies closest to them, either by merging or by the QSO radiation,
since there are fewer and fainter galaxies close to the QSOs. 
It would clearly be of interest to obtain high resolution deep imaging
of these QSO fields, to test this scenario and gather essential details
of the galaxy morphology. We thank the CFHT QSO team for making
the observations, and a referee for helping improve our presentation. 

\newpage
\centerline{References}

Bianchi et al, 2005, ApJ, 619, L27

Bianchi et al, 2007, ApJs, 173, 659

Bianchi, L. 2008, in "Space Astronomy: the UV window to the Universe"'
APSS, DOI: 10.1007/s10509-008-9761-3

Bianchi L., Hutchings J.B., Efremova B., Herald J.E., Bressan A., Martin C.,
2008, AJ (submitted)

Hutchings J.B., Frenette D., Hanisch R., Mo J., Dumont P.J., Redding D.C.,
Neff S.G., 2002, AJ, 123, 2936

Hutchings J.B., Proulx C., 2008, AJ, 135, 1692

Falomo R., Kotilainen J.K., Pagani C., Scarpa R., Treves A., 
2004, ApJ, 604, 495

Kotilainen J.K., Falomo R., Labita M., Treves A., Uslenghi M., 2007, ApJ,

Kukula M.J., Dunlop J.S., McLure R.J., Miller R., Percival W.J., Baum S.A.,
O'Dea C.P., 2001, MNRAS, 326, 1533

Lacy M., 2006, Astro=ph/0601255

Ridgway S.E., Heckman T.M., Calzetti D., Lehnert M., 2001, ApJ, 550, 122

Salviander S., Shields G.A., Gebhardt K., Bonning E.W., 2007, ApJ, 662, 131

Schlegel D.J., Finkbeiner D.P., Davis M., 1998, ApJ, 500, 525

Wold M., Lacy M., Lilje P.B., Serjeant S., 2001, astro=ph/0102046

\newpage
\centerline{Captions to Figures}

1. Distribution of FWHM for stars in the data, compared with those for 
QSOs, scaled to the same total.

2. Photometric calibration based on SDSS magnitudes and our mean fluxes
for our data on individual QSOs.

3. Looking for asymmetry-measure biases with image quality, brightness,
and redshift. The asymmetry measurements have no significant
correlation with these quantities.

4. Spatial distribution of all galaxies measured around the QSOs, 
and for randomly selected (control) positions far from the QSO.

5. Galaxy counts per unit area of sky, for the QSO and control fields.
Uniform density should yield linear plots. Linear fits (dotted lines) 
are shown for the control field galaxy counts. The upper plots include 
all galaxies to our flux limit, and the lower plots are those which had
asymmetry measurements on them. The ratio of total to asymmetry-measured
galaxies is about 3 (see text discussion).

6. Magnitude trends with projected distance from the QSOs, compared with 
field values. Points are all galaxies within the given radius from the QSO
or random control origin position.

7. GALEX and SDSS magnitudes for the sample QSOs (dots), with the median 
values connected by the thick line. The lighter lines represent medians 
from samples at different redshifts, with
bandpass bins shifted to match the z$\sim$1 sample observations. 

8. Mean normalised asymmetry for various objects as function of
magnitude. Filled dots are QSOs, open circles their companions, and
crosses the field galaxies. The QSO asymmetry values above the star
value, have been multiplied by a factor 10 to derive the values for
the host galazies, and compare with the other galaxies. 
The QSO host galaxy magnitudes have accordingly been increased by 2.5 
from the QSOs, for the same reason. To avoid crowding the diagram, star 
values are not plotted, but their median is shown at the edge of the box. 

9. Galaxy asymmetry with projected distance from the QSO (dots), and for 
field galaxies about random origin (circles). The dashed line shows the mean 
value for field galaxies and the two solid lines sketch in two possible
distributions for the QSO companions, as discussed in the text.

10. Distributions of asymmetry index for QSO companions (solid) and
field galaxies (dashed line). The distributions are different at the 
98\% confidence level.

\newpage

\small
\begin{deluxetable}{ccccccccccc}
\tablenum{1}
\tablecaption{Catalogued SDSS and GALEX properties of sample objects}
\tablehead{\colhead{RA} &\colhead{Dec} 
&\colhead{E$_{B-V}$}
&\colhead{FUV} &\colhead{NUV} &\colhead{u} &\colhead{g} &\colhead{r} 
&\colhead{i} &\colhead{z} &\colhead{Redshift} }
\startdata
00 32 59& 15 49 59& 0.056& 22.14& 20.76& 20.03& 19.78& 19.51& 19.60& 
19.23& 0.98\\
00 33 50& 15 46 14& 0.057& 23.37& 21.26& 20.25& 20.28& 20.01& 19.96&
20.13& 1.26\\
00 36 34& 14 35 46& 0.093& 20.39& 19.61& 19.66& 19.41& 19.07& 19.02& 
18.83& 1.01\\
00 41 59& 14 54 47& 0.091& 22.16& 20.52& 19.94& 19.79& 19.35& 19.21&
19.25& 1.18\\
00 44 41& 15 38 50& 0.048& 20.99& 19.94& 19.44& 19.23& 19.12& 19.17&
19.04& 0.93\\
00 45 46& 15 50 24& 0.050& 21.39& 19.59& 19.61& 19.54& 19.16& 19.18&
19.13& 1.16\\
00 46 14& 15 38 36& 0.056& 22.43& 21.06& 20.53& 20.46& 20.86& 19.97&
19.52& 1.16\\
00 46 36& 16 01 30& 0.059& 20.51& 19.52& 19.10& 18.91& 18.63& 18.60&
18.69& 1.12\\
00 54 44& 14 46 46& 0.054& 19.65& 18.36& 18.19& 17.93& 17.78& 17.80&
17.60& 0.91\\
01 17 44& 14 50 11& 0.056& 21.02& 20.05& 19.93& 19.63& 19.26& 19.10& 
19.12& 1.07\\
01 22 46& 14 32 03& 0.042& 21.60& 20.93& 20.69& 20.47& 20.32& 20.14&
20.41& 1.19\\
01 22 54& 14 51 03& 0.054& 19.65& 18.71& 18.68& 18.74& 18.42& 18.37&
 18.30& 1.23\\
01 23 00& 15 11 48& 0.059& 20.27& 19.34& 18.75& 18.69& 18.38& 18.33&
18.44& 1.28\\
01 23 06& 15 39 10& 0.083& 21.62& 20.35& 20.10& 19.96& 19.53& 19.36&
 19.56& 1.07\\
01 24 44& 13 26 42& 0.039& 21.29& 19.73& 19.01& 18.71& 18.47& 18.38&
 18.53& 1.26\\
01 47 27& 14 21 51& 0.050& 20.15& 18.93& 18.36& 18.27& 18.09& 18.09& 
18.00& 1.03\\
\enddata

\end{deluxetable}

\begin{deluxetable}{lccccccll}
\tablenum{2}
\tablecaption{Asymmetry measurements}
\tablehead{\colhead{Type} &\colhead{Mean} &\colhead{Mean g+i} 
&\colhead{Mean g}
&\colhead{Mean i} &\colhead{Number of}\\
&\colhead{asymm} &\colhead{asymm} &\colhead{asymm}
&\colhead{asymm} &\colhead{objects} }
\startdata
QSOs\tablenotemark{1} & 0.010\tablenotemark{2} & 0.008 & 0.008 & 0.012 
&16 \\
Stars\tablenotemark{3} &0.005 &&0.006 &0.004 &30\\
Galaxies (QSO Field) & 0.024 & 0.023 & 0.023 & 0.026 & 94 \\
Control Galaxies & 0.027 & 0.024 & 0.028 & 0.029 & 136 \\
QSO Galx $<$200px & 0.031 & 0.026 & 0.033 & 0.034 & 34 \\
Control Galx $<$200px & 0.023 & 0.022 & 0.025 & 0.024 & 33 \\
\enddata

\tablenotetext{1}{QSO host values (nucleus removed) $\sim$5x higher than
these values}
\tablenotetext{2}{does not include g+i measurements}
\tablenotetext{3}{median, matched to QSO magnitudes}
\end{deluxetable}

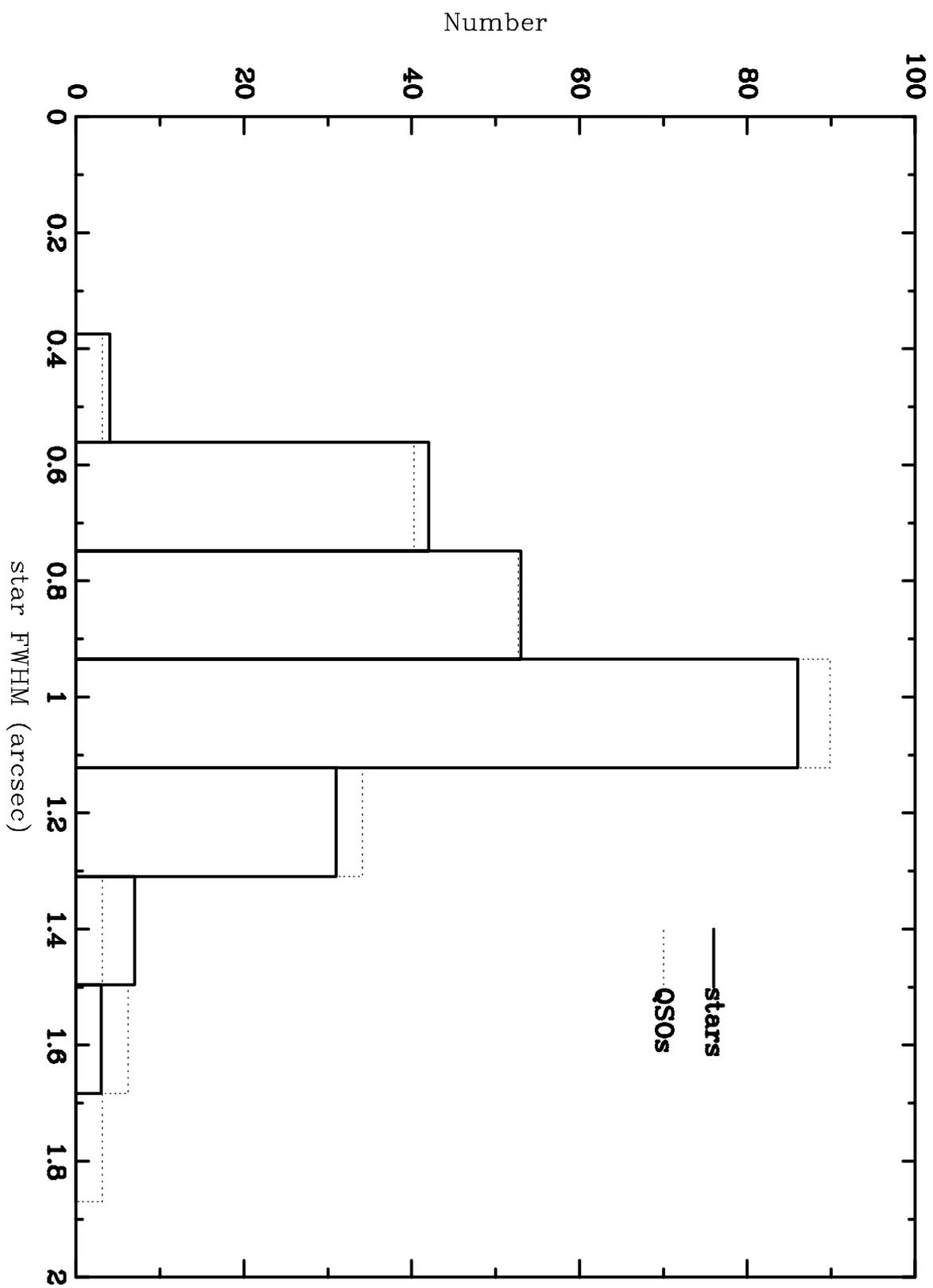
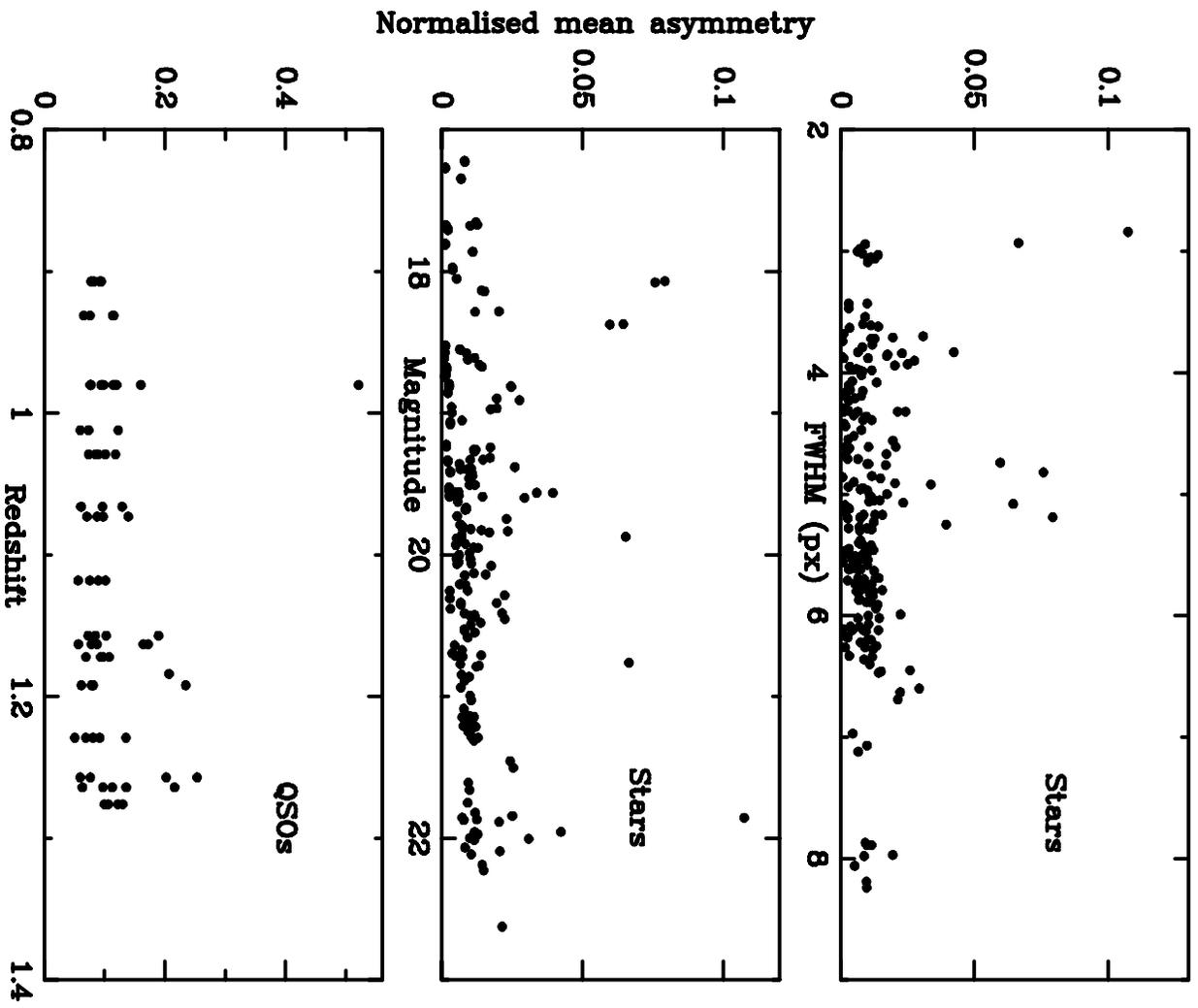
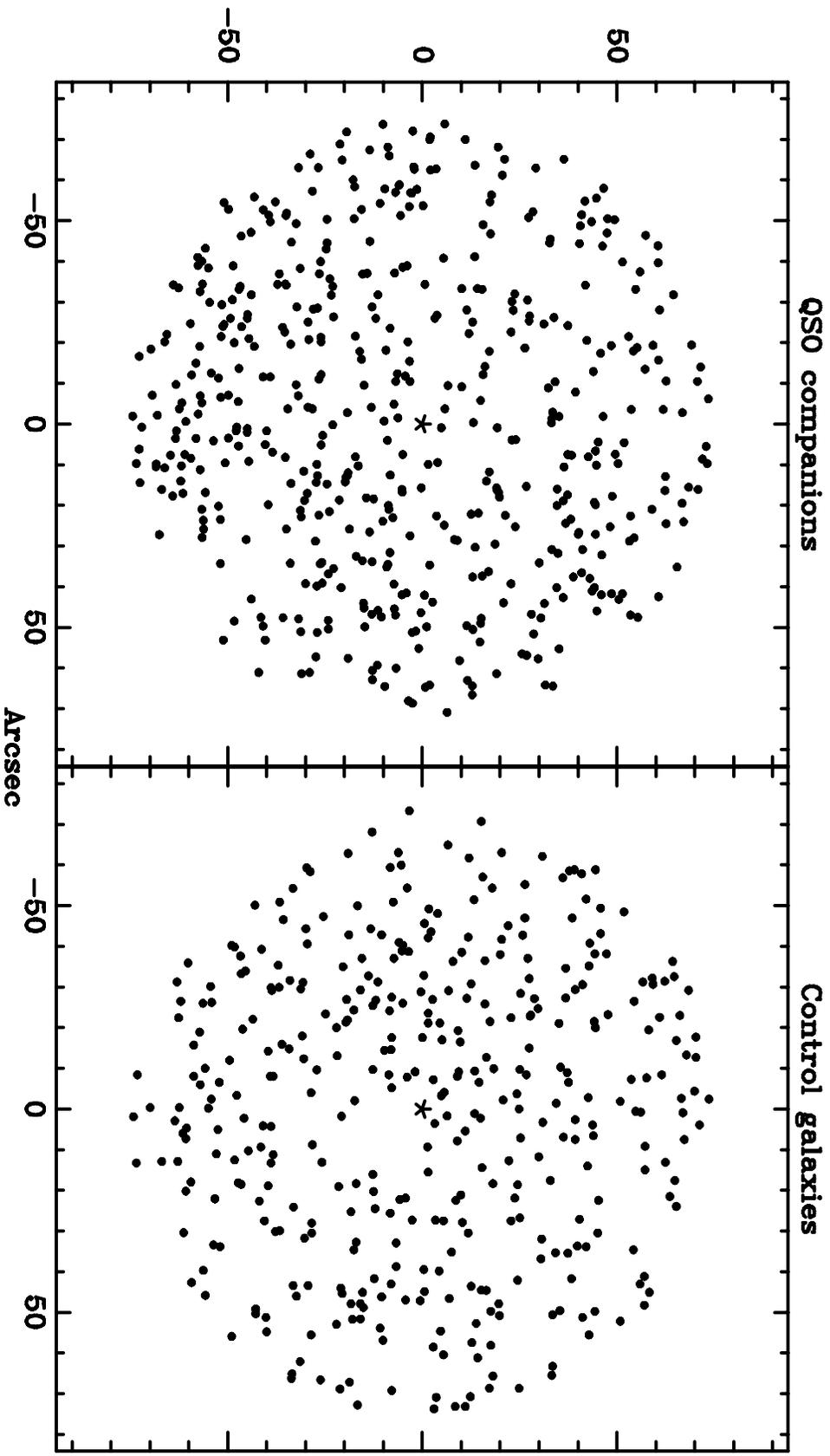
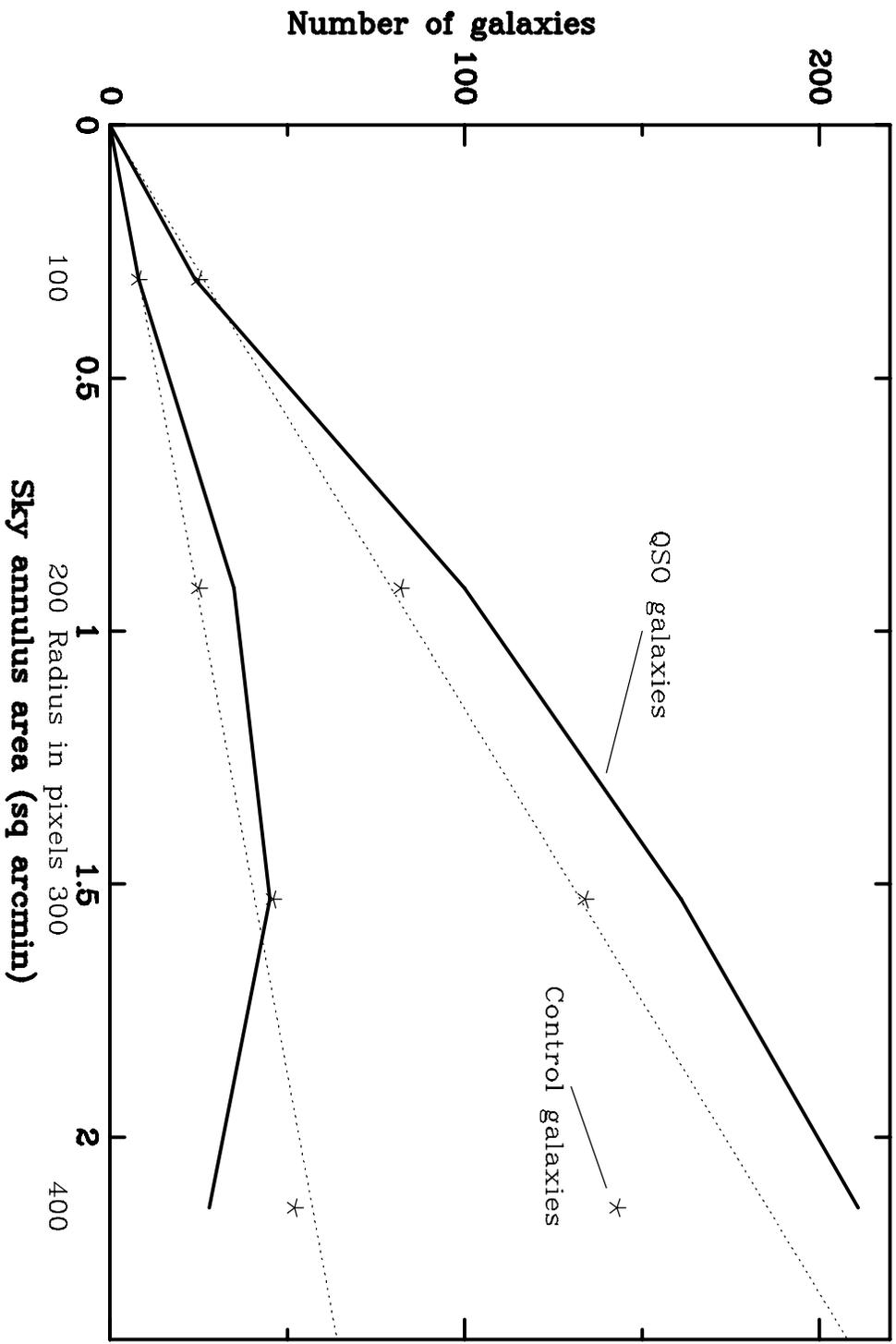
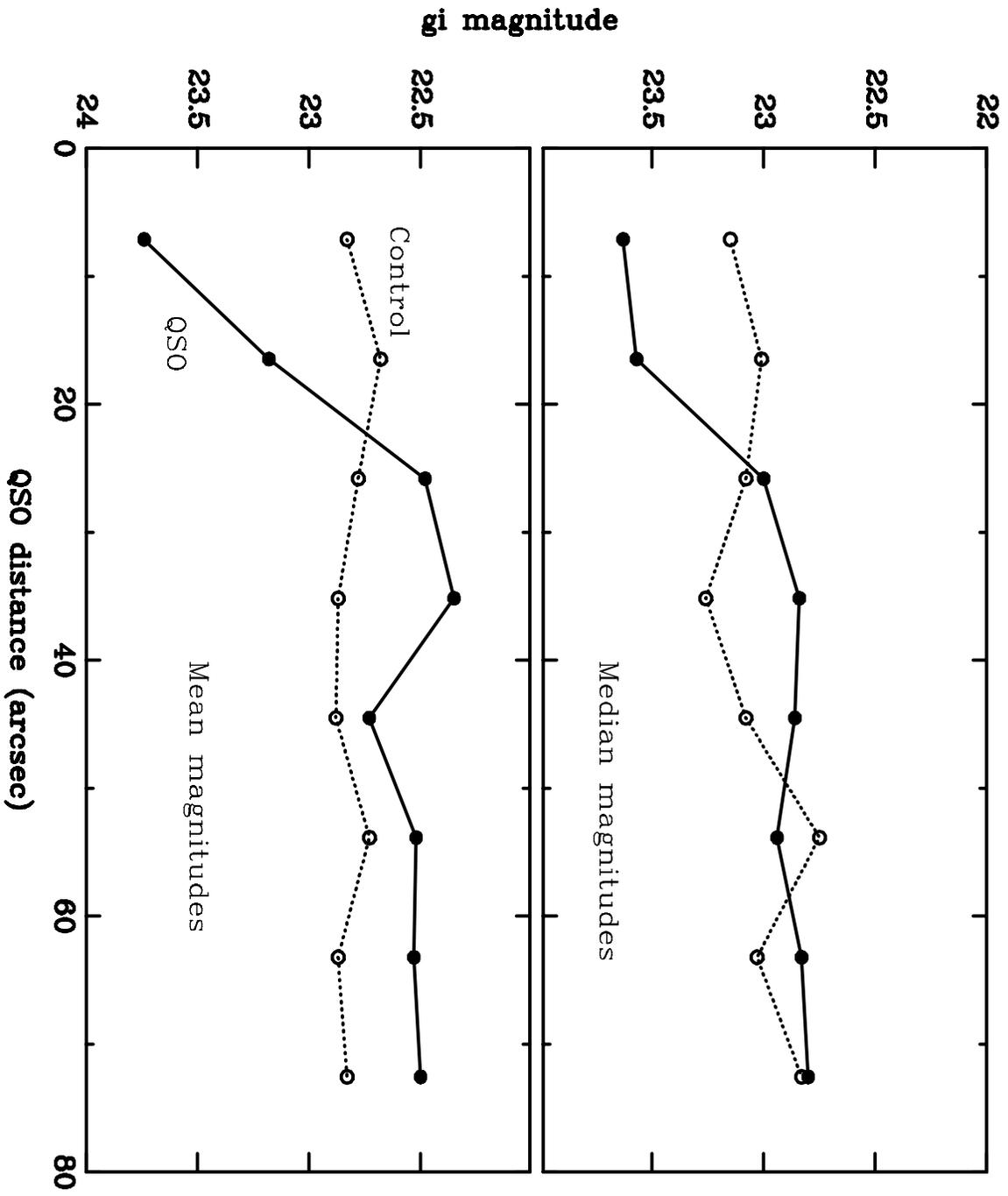
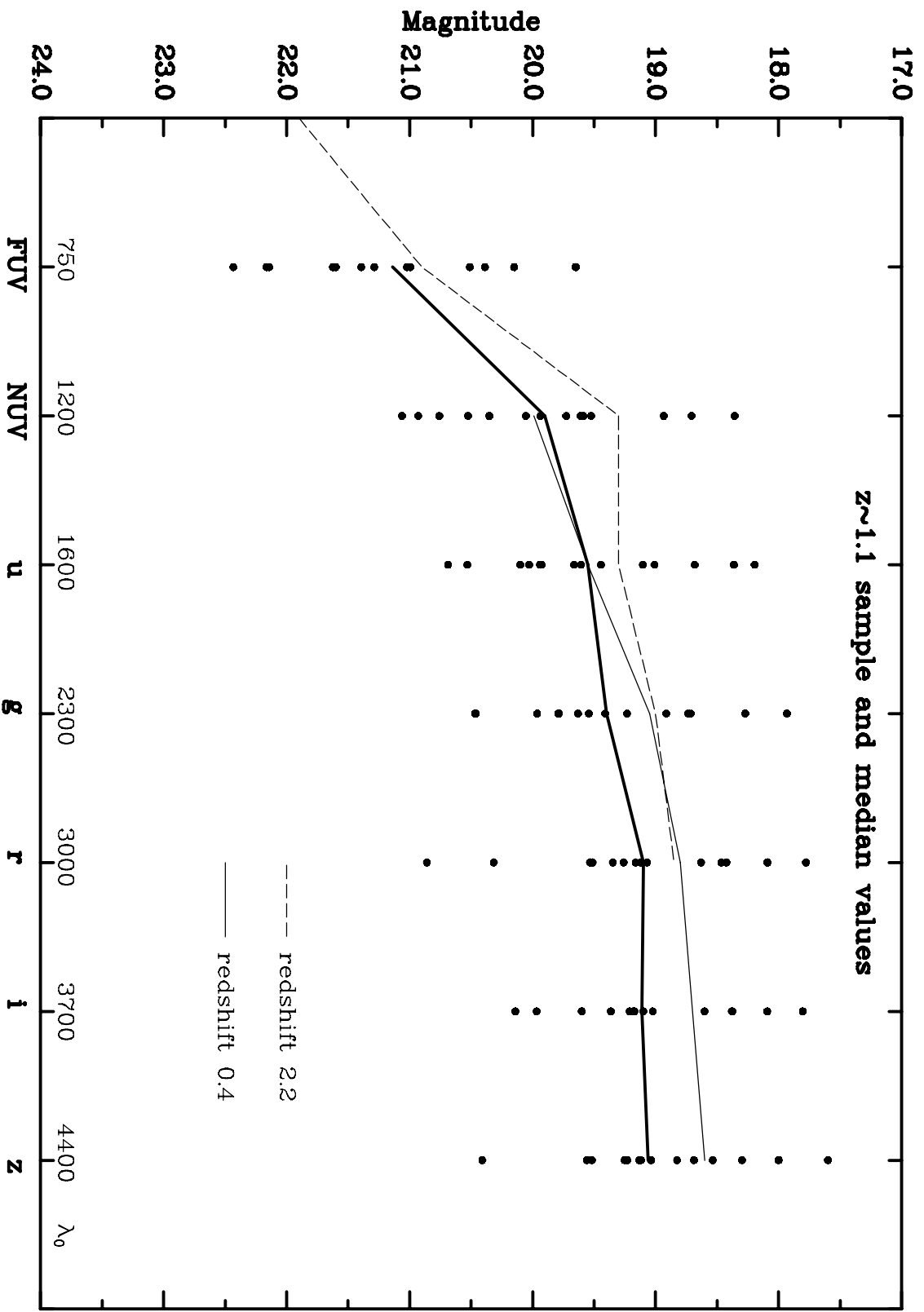
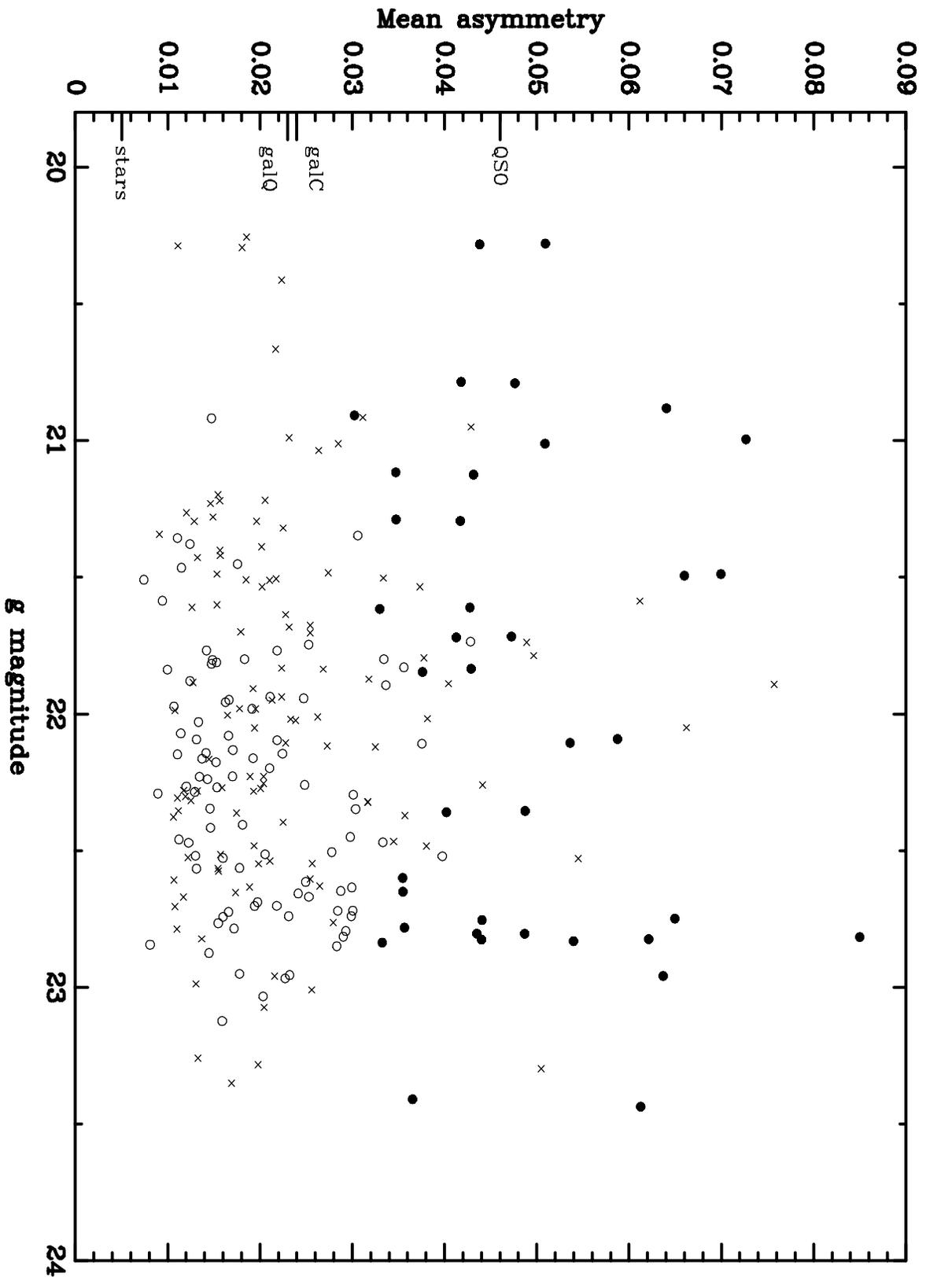
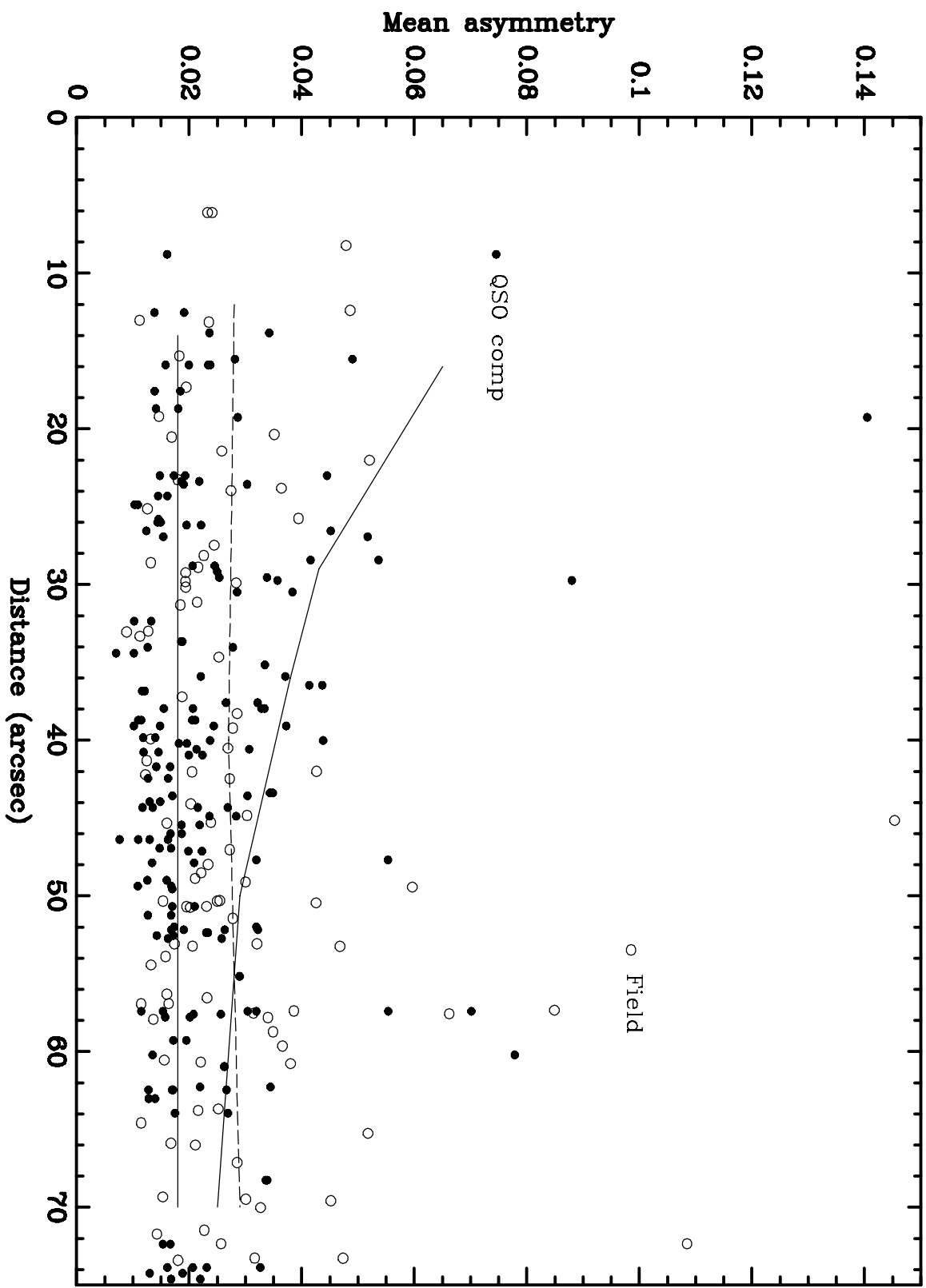
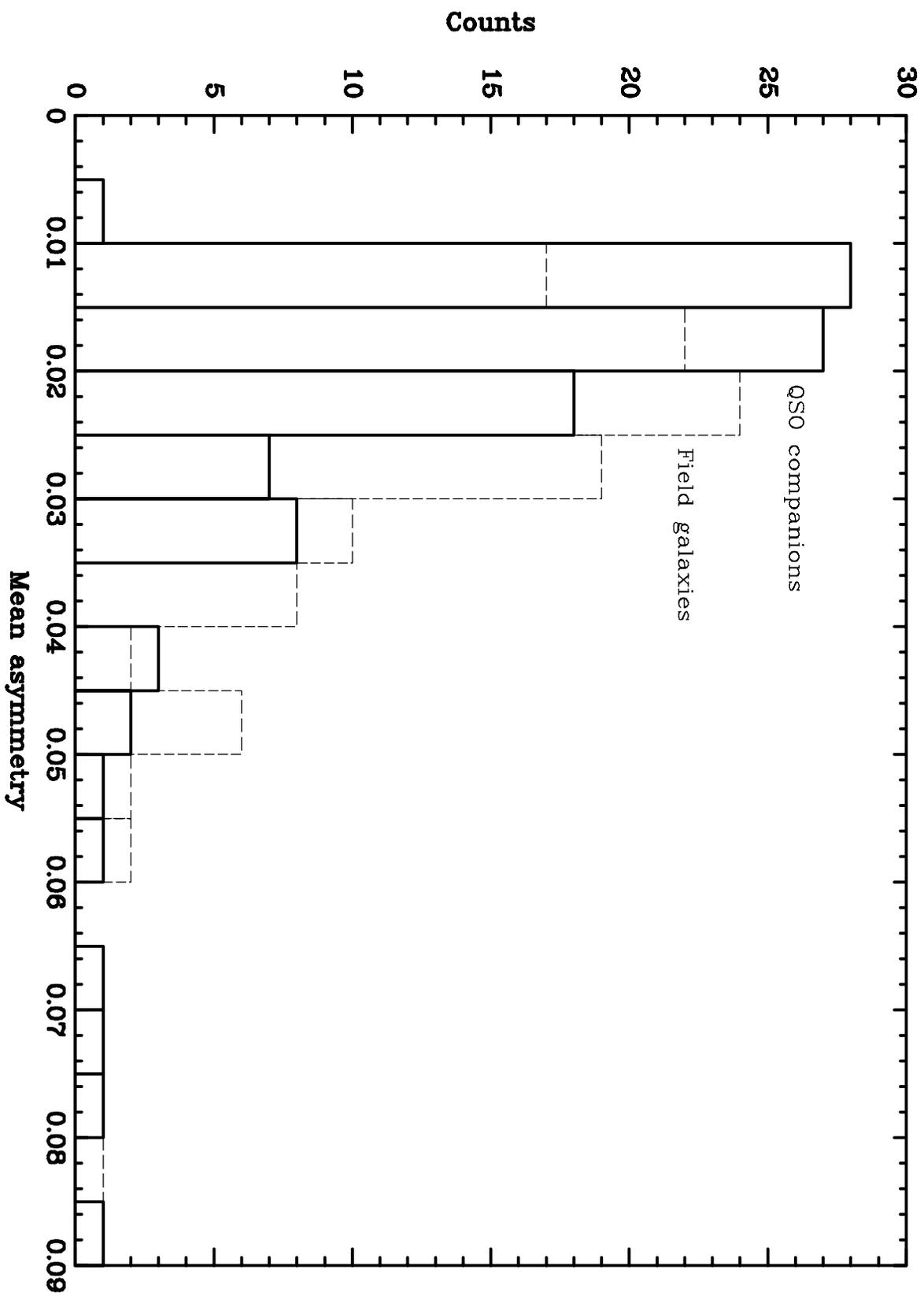
\end{document}